# Momentum transfer and foam production via breaking waves in hurricane conditions

Ephim Golbraikh[1] and Yuri M. Shtemler[2]

[1]*Department of Physics, Ben-Gurion University of the Negev, P.O. Box 653, Beer-Sheva 84105, Israel*

[2]*Department of Mechanical Engineering, Ben-Gurion University of the Negev, P.O. Box 653, Beer-Sheva 84105, Israel*

**Abstract**

Generated under hurricane conditions, a slip layer composed of foam, bubble emulsion, and spray determines the behavior of the surface drag with wind speed. This study enables us to estimate foam's contribution to this behavior. A logarithmic parametrization of surface drag is introduced, wherein the effective roughness length of the underlying surface is decomposed into three fractional roughness lengths. These correspond to the foam-free area (as determined by laboratory data, which includes the effects of spray and bubble emulsion) and ocean areas covered by whitecaps and streaks, each weighted by their respective coverage coefficients. A key concept of this approach is the use of well-established experimental bubble-size spectra produced by breaking surface waves to obtain the effective roughness length. This method provides a fair correlation of the logarithmic parametrization of surface drag against wind speed with a wide class of respective experimental data. Additionally, this approach estimates the hurricane's potential intensity, demonstrating reasonable agreement with experimental findings.

1. **Introduction**.

Surface wave breaking is crucial for ocean-atmosphere interaction. The breaking waves in hurricane conditions create an air-water layer consisting of foam, bubble emulsion, and spray, which functions as a slip layer. This layer significantly influences the transfer of momentum and heat between the ocean and atmosphere, with the impact varying with wind speed (Powell et al. 2003; Holthuijsen et al. 2012, hereafter referred to as HPP2012; Donelan 2018; Kudryavtsev and Makin 2011; Bye et al. 2014; Soloviev et al. 2014; Stemler et al. 2010; Golbraikh and Shtemler 2016, hereafter referred to as GS2016; Takagaki et al. 2016; Troitskaia at al. 2017; MacMahan 2017).

The surface transfer coefficients, $C_D$ and $C_K$, characterize the exchange of momentum and heat between the ocean and atmosphere. These coefficients change in response to the hurricane wind speed $U_{10}$ at a reference height of 10 meters. The process of momentum transfer from the atmosphere to the ocean under hurricane conditions has been the subject of extensive research (e.g. Powell et al. 2003; Donelan et al. 2004; Black et al. 2007; Jarosz et al. 2007; Shtemler et al. 2010; HPP2012; Troitskaya et al. 2019; Takagaki et al. 2012; Soloviev et al. 2014; GS2016; Golbraikh and Shtemler 2018, hereafter referred to as GS2018; Ermakova et al. 2023, and references therein). Also, the behavior of the enthalpy transfer coefficient against wind speed has been extensively examined (see e.g. Zhang et al. 2008; Bell et al. 2012; Richter et al. 2016; Jeong et al. 2012; Sergeev et al. 2017; Komori et al. 2018; Golbraikh and Shtemler 2020, hereafter referred to as GS2020). Numerous measurements in open-sea hurricane conditions demonstrate anomalous behavior of the momentum and enthalpy transfer coefficients, compared with laboratory conditions, at wind speed $U_{10}$. Specifically, $C_D$ monotonically increases with $U_{10}$, reaching saturation or even a reduction at certain speeds.

This phenomenon may be attributed to various physical mechanisms, such as Kelvin-Helmholtz instability at the air-sea interface, flow separation from the crests of breaking waves, and the influence of the slip layer. The role of foam within the slip layer and its impact on the momentum and enthalpy flux exchanges between the atmosphere and ocean have been discussed in (Newell, and Zakharov 1992; Powell et al. 2003; HPP2012; GS2016; GS2018; GS2020; Lan et al. 2022; Sergeev and Kandaurov 2022; Shtemler et al. 2010; Sroka, and Emanuel 2021; Sun et al. 2021; Troitskaya et al. 2019, and references therein). In particular, our earlier studies are based on a parametrization of the interface momentum and enthalpy fluxes. They involve splitting the effective roughness length into two weighted components for the foam-free- and the foam - covered areas. This approach explores that laboratory data for the interface fluxes are foam-free (see, for example, Donelan et al. 2004; Takagaki et al. 2012). It is assumed that the influence of the sliding through, for instance, spraying, and wave presence on interfacial flows is already included in their parameterization, and only the influence of foam requires separate consideration.

The following section presents a brief review, primarily focusing on our earlier studies into the impact of foam on momentum and enthalpy fluxes in hurricane conditions. Section 3 employs the roughness length splitting for the foam-covered surface, distinguishing between whitecap-covered and streak-covered areas. The resulting surface drag coefficient also estimates the hurricane's potential intensity. The conclusions are presented in Section 4.

**2. Foam influence on the momentum and enthalpy transfer in hurricanes**

As in many studies (e.g., Powell et al. 2003; Black et al. 2007; Powell 2007; HPP2012; Bryant, and Akbar 2016; Hsu et al. 2017), we assume that the average wind speed $U$ under hurricane conditions is well approximated by the logarithmic velocity profile in the boundary layer under conditions of neutral atmospheric stability:

$$U(Z) = \frac{U_*}{\varkappa} \ln\left(\frac{Z}{Z_D}\right). \tag{1}$$

$\varkappa = 0.4$ represents von Karman's constant; $U_*$ [ms$^{-1}$] is the friction velocity; $Z$ [m] is the current height above the sea surface; $Z_D$ [m] is the effective momentum roughness length over the total sea surface. The surface impulse flux $\tau = \rho C_D U_{10}^2$ ($\rho$ is the atmosphere density) is associated with the drag coefficient $C_D$ determined through the speed at a reference height of 10 [m]. Following (1) $C_D$ can be written as:

$$C_D(U_{10}) = \frac{U_*^2}{U_{10}^2} \equiv \varkappa^2 / \ln^2(10/Z_D). \tag{2}$$

For instance, the coefficient $C_D$ as a function of $U_{10}$ can be determined through direct measurements. Commonly, two experimental approaches are employed to investigate air-sea momentum transfer under open-sea conditions. The first is the 'top-down' approach, which involves recording the $U$ profile at considerably high altitudes, $H$. This measurement is typically approximated by the least squares line in the plane of the semi-log variables $\{U, \log H\}\}$. Subsequently, it is extrapolated down to sea level, where the average wind speed $U$ equals 0, to obtain the effective roughness length $Z_{eff}$. This methodology is supported by various studies, including Powell et al. 2003, Black *et al.* 2007, Powell 2007, HPP2012, Richter et al. 2016, among others. Nevertheless, the dispersion in the resulting dependence of $C_D$ on $U_{10}$ is considerable. The second approach is the 'bottom-up' method, which involves measuring the upper sea currents at the air-sea interface, as detailed in Jarosz et al. 2007. This technique focuses on the surface drag coefficient, which characterizes the sea-bottom stress in the linear drag law of the simplified momentum balance equation. Jarosz et al. 2007, and Bryant, and Akbar 2016 note that this method provides a reliable and relatively accurate determination of $C_d(U_{10})$, except for wind speeds less than $U_{10} = 30$ m s$^{-1}$. As a result, the $C_d(U_{10})$ values reported by Jarosz et al. 2007 are now commonly accepted as fiducial dependences (HPP2012, and Soloviev et al. 2014). All these experiments demonstrate saturation and a further decrease in the drag coefficient $C_d(U_{10})$ under hurricane conditions. However, these procedures do not provide an understanding of the physical processes leading to the non-monotonic behavior of $C_d(U_{10})$ under hurricane conditions.

A three-layer configuration, comprising the ocean, a foam layer, and the atmosphere, is introduced to model the momentum and energy transfers between the atmosphere and ocean (Newell and Zakharov 1992; Shtemler et al. 2010). In GS2016, the foam layer presence on the sea surface was hypothesized to describe the anomalous behavior of $C_D$ with $U_{10}$. Instead of relying on direct measurements, GS2016 determined the dependence of $Cd$ on $U_{10}$ by solving the inverse problem for roughness length $Z_D$. This approach involved two steps. Initially, a model was selected that accurately describes the behavior of $C_D(U_{10})$ in the hurricane range of $U_{10}$, but did not account for foam-covered areas as observed in laboratory experiments (i.e. $C_D(U_{10}) = C_{Dw}(U_{10})$, see Donelan et al. 2004; and Takagaki et al. 2012). Subsequently, $Z_D$ was presented as a splitting of two distinct roughness lengths: $Z_{D,w}$ for the water (foam-free)-covered area and $Z_{D,f}$ for the foam-covered area, each weighted by their respective fractional coverage coefficients, $\alpha_w(U_{10})$ and $\alpha_f(U_{10})$:

$$Z_D = \alpha_w Z_{D,w} + \alpha_f Z_{D,f}, \qquad (\alpha_w + \alpha_f = 1), \qquad (3)$$

where $\alpha_f = S_f/S$ and $\alpha_w = S_w/S$, $S_w$ represents the foam-free area, $S_f$ is the foam-covered area, and $S = S_w + S_f$ denotes the total sea surface area. In this framework, $Z_{D,w}$ and $Z_{D,f}$ are determined by solving the inverse problem (2) for $Z_{D,w}$ and $Z_{D,f}$ using available empirical data for $C_{D,w}(U_{10})$ and $C_{D,f}(U_{10})$, respectively. The dependence of $C_{D,w}(U_{10})$ (see Fig. 1) can be established through either empirical correlations or laboratory experiments (e.g., Large and Ponds 1981; Donelan et al. 2004; Komori et al. 2018). The dependence of foam coverage coefficient $\alpha_f$ against $U_{10}$ can be obtained from observational open-sea data in a wide range of $U_{10}$ values, as demonstrated by HPP2012. The data measured under hurricane conditions were approximated in HPP2012 as:

$$\alpha_f = \gamma \tanh\left[\alpha \exp\left(\beta \frac{U_{10}}{U_{10}^{(S)}}\right)\right], \tag{4}$$

where $\alpha = 0.00255$, $\gamma = 0.98$, the formula for $\beta = 0.166 U_{10}^{(S)} \approx 8$ was modified in GS2016 by incorporating a saturation velocity, set at $U_{10}^{(S)} \approx 48 \ [ms^{-1}]$.

Thus, the fractional roughness lengths, $Z_{D,w}$ and $Z_{D,f}$, are determined by solving the inverse problem (2) with available empirical data for $C_{D,w}(U_{10})$ and $C_{D,f}(U_{10})$, along with the foam coverage coefficient $\alpha_f(U_{10})$. Subsequently, the value of $C_D(U_{10})$ can be found using equations (2) and (3). The dependence $C_{D,w}(U_{10})$ is now well established. The parameter $C_{D,w}(U_{10})$ as determined in laboratory studies (Donelan et al. 2004; Takagaki et al. 2012) correlates well with phenomenological open-sea model of Large and Ponds 1981, as modified in GS2018 (see Fig. 1). The Large -Ponds model somewhat overestimated the data from Donelan et al. 2004 but showed good agreement with the data in Takagaki et al. 2012. The dependence $Z_{D,f}(U_{10})$ has been evaluated through various methods. In GS2016, the fractional roughness length $Z_{D,f}$ was estimated as an effective constant radius of foam bubbles, $R_f$, determined by the value $C_D$ obtained for an area completely covered by foam at $U_{10} > 50 \ [ms^{-1}]$ using experimental data from Powell et al. 2003; Edson et al. 2007; Black et al. 2007; and Jarosz et al. 2007. This yields the estimate for $Z_{D,f} \approx R_f = 0.3 \ [mm]$ independent of the velocity $U_{10}$. Therefore, the approach adopted in GS2016 provides a reasonable estimation for $C_D(U_{10})$. Thus, the key idea used in GS2016 is to equate the roughness parameter in the logarithmic velocity profile with the geometric parameter of the effective radius of the foam bubble. Although this estimation is strictly valid only at $U_{10} > 50 \ [ms^{-1}]$, as will be shown later (see Fig. 5), it is justified by a relatively minor contribution of whitecaps to $C_D(U_{10})$ in the most range of interest for $U_{10}$. A limitation of this approach is that approximation (4) for $\alpha_f(U_{10})$, as seen from the data in HPP2012, depends on the fixed value of the saturation velocity $U_{10}^{(S)} = 48 \ [ms^{-1}]$, while reliable data for $\alpha_f(U_{10})$ are only available for $U_{10} \leq 48 \ [ms^{-1}]$. Hence, the actual value of $U_{10}^{(S)}$ is rather a parameter that may vary with specific hurricane conditions.

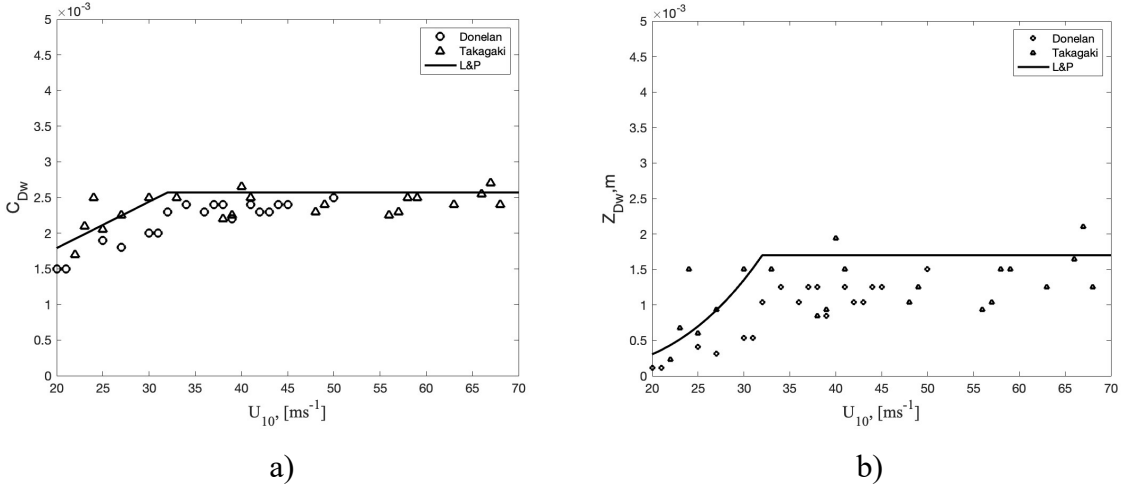

a)    b)

Fig. 1: The dependence a) $C_{Dw}$ vs $U_{10}$ and b) $Z_{Dw}$ vs $U_{10}$ based on laboratory experimental data from (Donelan et al. 2004; Takagaki et al. 2012) and on phenomenological correlations from (Large and Ponds 1981, as modified in GS2018). Stars represent data from Donelan et al. 2004, triangles represent data from Takagaki et al. 2012, and the solid line depicts the correlation from Large and Ponds 1981, as modified in GS2018.

GS2018 examined the actual dependence of $R_f$ on wind speed $U_{10}$ under hurricane conditions. They derived a phenomenological dependence of $R_f \approx Z_{D,f}(U_{10})$ from relation (3), solving the inverse problem for $Z_D$ using experimental data for $C_D(U_{10})$ provided by Jarosz et al. 2007, which was considered sufficiently reliable. Unfortunately, the significant scatter in the measurement data for $C_D(U_{10})$ under hurricane conditions led to an even greater scatter in $Z_{D,f}$ when solving the inverse problem, noticeably more than depicted in Fig. 1 for the foam-free case. The corresponding values of $R_f \approx Z_{D,f}(U_{10})$ could reach several millimeters (as shown in Figure 2 in GS2018) and were somewhat overestimated. However, this estimation is checked against the Hinze scale, $r_H$, which predicts the maximum stable bubble size before they break up into smaller bubbles. According to this scale, the size does not exceed the range of $10^{-3}$ to $1.8 \cdot 10^{-3}$ meters (Hinze 1995, Deane and Stokes 2002, Soloviev and Lukas 2006, and Blenkinsopp et al. 2010). Measurement data for the momentum and enthalpy transfer coefficients, $C_D(U_{10})$ and $C_K(U_{10})$ at wind speeds, $U_{10}$, greater than $50 - 60 \ [ms^{-1}]$ are very scarce (Powell et al. 2003; Bell et al. 2012; Richter et al. 2016; Sroka, and Emanuel 2021).

These investigations reveal a local minimum ($C_D(U_{10}) \approx 0.001$) in these coefficients, followed by an increase at wind speeds exceeding 60 [$ms^{-1}$]. This increase at very high wind speeds may be attributed to the partial destruction of the foam layer. Due to the lack of data on $\alpha_f(U_{10})$ at such high wind speeds, the phenomenological formula (4) for the foam coverage coefficient $\alpha_f(U_{10})$ was slightly adjusted in GS2020 by integrating a linear correction:

$$\alpha_f(U_{10}) = \begin{cases} \gamma \tanh\left[\alpha \exp\left(\beta \frac{U_{10}}{U_{10}^{(S)}}\right)\right] & \text{for } U_{10} \leq 52 [ms^{-1}] \\ \gamma \tanh\left[\alpha \exp\left(\beta \frac{U_{10}}{U_{10}^{(S)}}\right)\right] - \delta U_{10} + \varepsilon & \text{for } U_{10} > 52 [ms^{-1}], \end{cases} \quad (5)$$

where $\alpha$, $\beta$, $\gamma$, are as defined in formula (4) with $\delta = 0.004347$ and $\varepsilon = 0.22$ chosen in GS2020 to align the modified models of $C_D$ and $C_K$ with experimental data for high wind speeds exceeding 50 - 60 [$ms^{-1}$].

GS2020 evaluated the impact of the foam slipping layer, sandwiched between the atmosphere and ocean, on the enthalpy transfer coefficient $C_K$. Similar to the previously discussed model for the momentum transfer coefficient $C_D$ in GS2016, $C_K$ in GS2020 is modeled as:

$$C_K = (1 - \alpha_f) C_{K,w} + \alpha_f C_{K,f}. \quad (6)$$

The enthalpy transfer coefficients over the water-covered sea surface, $C_{K,w}$, and foam-covered, $C_{K,f}$, portions of the sea surface are defined respectively as:

$$C_{K,w}(U_{10}) = \begin{cases} 1.39 \cdot 10^{-3} & \text{for } U_{10} \leq 33.6 \ [ms^{-1}], \\ 6.51 \cdot 10^{-5} U_{10} - 7.99 \cdot 10^{-4} & \text{for } U_{10} \geq 33.6 \ [ms^{-1}], \end{cases} \quad (7)$$

$$C_{K,f}(U_{10}) \approx 0.00085. \quad (8)$$

Here $C_{K,w}(U_{10})$ was established based on laboratory experiments by Komori et al. 2018, whereas $C_{K,f}(U_{10})$ was modelled using experimental data from Bell et al. 2012, and Richter et al. 2016, indicating that $C_{K,f}(U_{10})$ does not vary with $U_{10}$. The physical parameter that characterizes the hurricane potential intensity, $C_K(U_{10})/C_D(U_{10},)$ was found in GS2020 to be beyond the experimentally observed range of values.

## 3. The momentum transfer and foam production via breaking waves in hurricanes

This study addressed to avoid the significant errors associated with scattering in the foam fractional roughness length, $Z_{D,f}$, often observed in solutions to the inverse problem under hurricane conditions. Besides, the modeling aims to better align with the acceptable theoretical and experimental estimation for the ratio $C_K(U_{10})/C_D(U_{10})$. In this study, our goal is to avoid the significant, by using for modeling the characteristics of sea foam bubbles produced due to the surface waves breaking under hurricane conditions.

The capture of atmospheric air by the water during wave breaking results in a foam-covered ocean surface under hurricane conditions (e.g. Soloviev and Lukas 2006). Extensive experimental and theoretical research has identified the bubble-size distribution function $\Phi_f(r)$ that describes the population of foam bubbles produced by breaking waves (Bowyer 2001; Ding et al. 2002; Deane and Stokes 2002; Han and Yuan 2007; Deike et al. 2016; Mostert et al. 2022). The foam-covered ocean surface comprises whitecaps and streaks (HPP2012), are characterized by distinct bubble-size distribution functions, $\Phi_{wc}(r)$ and $\Phi_{str}(r)$, respectively. Importantly, both $\Phi_{wc}(r)$ and $\Phi_{str}(r)$ are independent of the wind speed, $U_{10}$ (e.g. Mostert et al. 2022). These distribution functions determine the effective characteristic radii of foam bubbles in whitecaps and streaks, $R_{wc}$ and $R_{str}$, which are also independent of $U_{10}$. Hence, the fractional foam roughness, $Z_{D,f}$, can be naturally split into two sub-fractional roughness lengths, $Z_{D,wc}$ and $Z_{D,str}$ for the whitecap- covered area and for streak-covered ocean area. These are weighted by the fractional coverage coefficients, $\alpha_{wc}(U_{10})$ and $\alpha_{str}(U_{10})$, respectively:

$$\alpha_f Z_{D,f} = \alpha_{wc} Z_{D,wc} + \alpha_{str} Z_{D,str}, \quad (\alpha_f = \alpha_{wc} + \alpha_{str}). \tag{9}$$

In this context, $Z_{D,wc}$ is equal to $R_{wc}$, and $Z_{D,str}$, is equal to $R_{str}$, where $R_{wc}$, and $R_{str}$ are the corresponding effective radiuses of the bubbles. Furthermore, $\alpha_{wc} = \frac{S_{wc}}{S}$, $\alpha_{str} = \frac{S_{str}}{S}$, where $S$ is the total surface area ($S = S_w + S_{wc} + S_{str}$), with $S_w$, $S_{wc}$ and $S_{str}$ indicating the water-covered, whitecap- covered and streak- covered areas, respectively.

HPP2012 indicates that the coefficients $\alpha_{wc}$ and $\alpha_{str}$ are influenced by the wind speed, $U_{10}$. The ocean surface covered by foam created by wave breaking increases with $U_{10}$, and so does the amplitude of waves (Hwang and Walsh 2018, and Lin and Oui 2019), resulting in a higher volume of air entrapped during the breaking process (Ding 2002). This mechanism leads to an increasing foam formation, expanding its coverage of the ocean surface. Zhang and Oey 2019 found that the maximum wave amplitude reaches saturation at wind speeds, $U_{10}$, of roughly $50 - 55 \ [ms^{-1}]$. At low wind speeds, the values of $\alpha_{wc}$ and $\alpha_{str}$ are nearly zero. The value of $\alpha_{wc}$ increases with $U_{10}$, reaching a peak at approximately 25 - 30 $[ms^{-1}]$ and then stabilizing as $U_{10}$ continues to rise. Conversely, $\alpha_{str}$ continues to rise with the increasing volume of air trapped during surface-wave breaking, achieving its maximum at $U_{10} \approx 50\ [ms^{-1}]$, which aligns with the saturation point of wave amplitude.

A crucial aspect of the proposed approach is that the roughness lengths, $Z_{D,ws}$ and $Z_{D,str}$, are considered equal to the effective radii of foam bubbles in whitecaps, $R_{wc}$, and streaks, $R_{str}$. These lengths are universal for any hurricane, independent of wind speed $U_{10}$, and entirely governed by the surface-wave breaking. However, these lengths evidently depend on the water properties like salinity, temperature, and biological content, which can differ in hurricanes. Besides, the effective foam roughness length, $Z_{D,f}$, depends on the experimental data for specific hurricanes through the fractional coverage coefficients of whitecaps $\alpha_{wc}(U_{10})$ and streaks $\alpha_{str}(U_{10})$, which generally vary with $U_{10}$, as noted by HPP2012.

The effective momentum roughness length, $Z_D$, can be ad hoc split into three roughness lengths: $Z_{D,w}$ for water-covered surface, $Z_{D,wc}$ for whitecap-covered surface, and $Z_{D,str}$ for streak-covered ocean surface. These are weighted by their respective fractional coverage coefficients, $\alpha_w$, $\alpha_{wc}$ and $\alpha_{str}$:

$$Z_D = \alpha_w Z_{D,w} + \alpha_f Z_{D,f}, \qquad (\alpha_w + \alpha_f = 1,\ \alpha_f = \alpha_{wc} + \alpha_{str}), \tag{10}$$

where

$$\alpha_f Z_{D,f} = \alpha_{wc} Z_{D,wc} + \alpha_{str} Z_{D,str}.$$

Summarizing, note that the foam roughness, $Z_{D,f}(U_{10})$, can be determined from Eq. (9) without solving the inverse problem (2)-(3) for $Z_{D,f}(U_{10})$ through $C_D(U_{10})$.

Whitecaps originate from comparatively large bubbles generated by hurricane-driven breaking waves and remain close to them due to the short lifespan of the bubbles. As long-lived streaks, with bubble sizes exceeding the Hinze scale, tend to disappear (Deane and Stokes 2002; Deike et al. 2016; Mostert et al. 2022), the effective bubble radius in streaks is significantly smaller compared to whitecaps. As was found that the fractional whitecaps coverage coefficient, $\alpha_{wc}$, increases with $U_{10}$, and saturates to a maximum value of approximately $0.05 - 0.1$ at greater than $25\ [ms^{-1}]$ (see Anguelova and Webster 2006; HPP2012; Brumer et al. 2017; Derakhti et al. 2023 and references therein). Based on data from HPP2012, the present study, focusing on hurricane wind speeds exceeding $25\ [ms^{-1}]$, assumes the following simple correlation:

$$\alpha_{wc}(U_{10}) \equiv const = 0.05. \tag{11}$$

The coverage coefficient, $\alpha_f(U_{10})$, has received comparatively less attention in experimental studies. It is evident that $\alpha_f(U_{10})$ increases with $U_{10}$ and approaches a maximum value in (5) as shown in Figure 2. In GS2018, the wind parameter $U_{10}^{(S)}$ was introduced in approximating formula by HPP2012 for convenience its further employment. Its value was set at 48 $[ms^{-1}]$, even though reliable data for the foam-coverage coefficient was not available for $U_{10}$ greater than 50 m/s. In addition, the measurements of the $\alpha_f(U_{10})$ were also carried out in RAINEX and CBLAST projects (Melville et al. 2010, El-Nimri et a 2010). Although their report does not indicate the error bars, the dependence (5), proposed in HPP2012, describes the data well with $U_{10}^{(S)} = 56\ [ms^{-1}]$. However, the value of $U_{10}^{(S)}$ may differ for different hurricanes or even for different hurricane zones. It can be addressed to the water properties (salinity, biological inclusions, etc.) in the ocean under the conditions under consideration.

Besides, Fig. 2 represents the distribution $\alpha_f(U_{10})$, corresponding to the value $U_{10}^{(S)} = 56\ [ms^{-1}]$ in correlation (5), which minimizes the diversity between the current model and the fiducial experimental dependence $C_D(U_{10})$ by Jarosz et al. 2007 (see Fig. 5 below). It should be noted that changing the value of $\alpha_{wc}(U_{10})$ has little effect on the resulting dependence $C_D(U_{10})$. Thus, the parameter, $U_{10}^{(S)}$, serving the best fitting of measurements data for $\alpha_f(U_{10})$ can be treated as a physical parameter that varies with specific hurricane conditions which correlate with variations in experimental data for $C_{D,f}(U_{10})$.

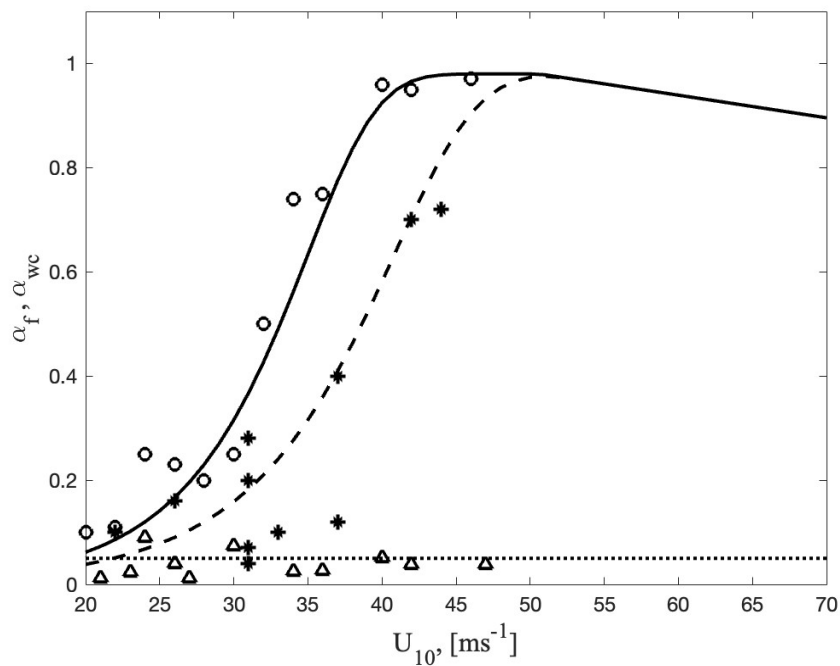

Fig. 2: Coefficients for total streak- and whitecap-coverage areas $\alpha_f = \alpha_{str} + \alpha_{wc}$, and for whitecaps $\alpha_{wc}$ vs $U_{10}$.

The solid and dotted lines represent the coverage coefficients, $\alpha_f(U_{10})$ (Eq. (5) with $U_{10}^{(S)} = 48\ [ms^{-1}]$) and $\alpha_{wc}(U_{10}) = 0.05$ (Eq. (11)) approximating experimental data by HPP2012 (circles and triangles correspond to experimental data for $\alpha_f(U_{10})$ and $\alpha_{wc}(U_{10})$, respectively). The dashed line represents $\alpha_f(U_{10})$ approximating experimental data by Melville et al. 2010 (Eq. (5) with $U_{10}^{(S)} = 56\ [ms^{-1}]$ (stars correspond to experimental points).

The roughness lengths of whitecaps, $Z_{D,wc}$, and streaks, $Z_{D,str}$, respectively, are determined using the bubble-size distribution functions $\Phi_{wc}(r)$ and $\Phi_{str}(r)$. These functions represent the number of foam bubbles of a given size as a function of their radius $r$. Breaking waves create whitecaps at the winds crests that follow the moving waves. As wind speeds rise, whitecaps remain constant and combined with increasingly dominated wind-aligned streaks of foam (e.g. HPP2012). This results in the independence of $\Phi_{wc}$ and $\Phi_{str}$ from the wind speed $U_{10}$. Experiments and numerical modeling show that $r = r_0 \approx 0.1mm$ corresponds to the maximum of $\Phi_{wc}(r)$. Distribution functions normalized to their maximal values will be further considered (see Fig. 3). Jumps observed in the slopes of the experimental spectra correspond to the transition between the sub-Hinze and super-Hinze regimes. Specifically, $\Phi_{wc}(r) \propto r^{-3/2}$ for $r < r_H$ and $\Phi_{wc}(r) \propto r^{-10/3}$ for $r > r_H$. These results are supported in numerous researches (Garrett et al. 2000; Bowyer 2001; Deane and Stokes 2002; Callaghan et al. 2014; Deike et al. 2016; Mostert et al. 2022 and references therein). Throughout this study, the value $r_H \approx 1.5\ [mm]$ is adopted for the Hinze scale (Deane and Stokes 2002, see Fig. 3).

Similarly to $\Phi_{wc}$, $\Phi_{str}$ exhibits a maximum value at $r = r_0$. Large bubbles ($r > 0.7\ [mm]$) are practically absent in streaks, as evidenced in (Carmill and Ming 1993; Bowyer 2001; Terrill et al. 2001; Leifer et al. 2003; Han and Yuan 2007; Deike et al. 2016, and references therein). Smaller bubbles have a longer lifespan than larger ones due to their stabilization by surface tension. They disintegrate slowly, especially when their radius is less than the Hinze scale. Simultaneously, whitecaps transform into streaks as bubbles with smaller radii persist for an extended period. The streak spectrum inhibits two regimes from the whitecap bubble-size distribution, $\Phi_{str} \propto r^{-n}$, in the range $r_0 \leq r \leq r_H$ (with $3 \leq n \leq 4$ and an average value of $n = 3.5$ adopted in this study), and $\Phi_{str} \propto r^{-5}$ for $r > r_H$. This steep distribution function decreases rapidly with increasing $r$ for $r > r_H$. Consequently, the spectrum $\Phi_{str}(r) \propto r^{-3.5}$ can be applied across the entire range of $r$, ensuring accurate calculations of the effective bubble radius in the forthcoming relation (12).

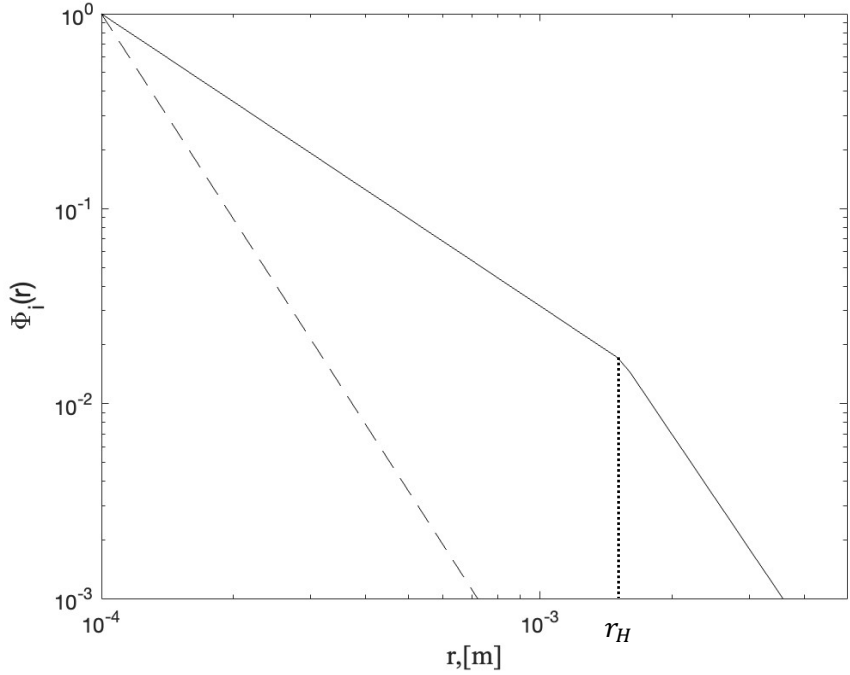

Fig.3: Normalized size distribution functions $\Phi_i(r) \sim r^{-n}$: for whitecaps represented by a solid line ($i = wc, n = \frac{3}{2}$ at $r_0 \leq r \leq 1.5 \cdot 10^{-3}[m]$, and $n = \frac{10}{3}$ at $r > 1.5 \cdot 10^{-3}[m]$), and for streaks represented by a dashed line ($i = str, n = 3.5$ at $r > r_0$). These functions are adopted from the studies by Garrett et al. 2000; Bowyer 2001; Terrill et al. 2001; Deane et al. 2002; Leifer et al. 2003; Blenkinsopp et al. 2010; Callaghan et al. 2014; Deike et al. 2016; Mostert et al. 2022.

In our papers GS2016 and GS2018, the roughness of a foam-covered surface is determined by the effective radius of foam bubbles on the ocean surface. Foam bubbles are distributed differently in size for the whitecap-covered and streak-covered areas. Therefore, in this study the effective radius $R_i$ of the foam bubbles is expressed using the bubble-size distribution function, $\Phi_i(r)$, calculated separately for the whitecap-covered ($i = wc$) and streak- covered ($i = str$) areas:

$$R_i = \int_{r_0}^{\infty} r\, \Phi_i(r) dr \Big/ \int_{r_0}^{\infty} \Phi_i(r) dr, \ (i = wc, str). \tag{12}$$

The lower limits in integrals (12) are equal to $r_0$ since according to experimental data for the spectra contribution to $R_i$ from the region between 0 to $r_0$, can be neglected (Bowyer 2001; Deane et al. 2002; Leifer et al. 2006; Blenkinsopp et al. 2010; Deike et al. 2016).

Utilizing the spectra of $\Phi_{wc}(r)$ and $\Phi_{str}(r)$ presented in Fig. 3, it can be inferred that the effective radii $R$ in the zones covered by whitecaps and streaks are:

$$Z_{D,wc} = R_{wc} \approx 0.0005 \, [m]. \text{ and } Z_{D,str} = R_{str} \approx 0.00015 \, [m]. \qquad (13)$$

The roughness of the foam-covered area $Z_f(U_{10})$ is depicted in Fig. 4 for $U_{10}^{(S)} = 48 \, [ms^{-1}]$ and $U_{10}^{(S)} = 56 \, [ms^{-1}]$. Since $R_{wc}$ is approximately 3 times greater than $R_{str}$ at velocities below $25 - 27 \, [ms^{-1}]$, the roughness of the foam cover is determined mainly by the whitecaps. For wind speeds, $U_{10}$, greater than $35 \, ms^{-1}$, the primary contribution to $Z_f$ comes from streaks, which predominantly cover the surface.

The roughness dependence $Z_D = Z_f(U_{10})$ is completely described by relations (10) - (11), where the effective roughness of streak- and whitecap-covered areas given by (13) with the help of numerous and reliable experimental spectra (12).

Figure 5 presents surface drag coefficient $C_D(U_{10})$ calculated with the current model for two sets of the input experimental profiles $\alpha_f(U_{10})$ adopted from HPP2012 and Melville et al. 2010 (see Fig.2), and for $\alpha_{wc} = 0.05$, the same for both cases. In Figure 5 $C_D(U_{10})$ calculated with the current model compared with fiducial experimental profile $C_D(U_{10})$ by Jarosz et al. 2007. It is seen that α_f (U_10) adopted from Melville et al. 2010 correspond to saturation wind speed $U_{10}^{(S)} = 56 \, [ms^{-1}]$,, , minimizes the diversity between the model and fiducial (by Jarosz et al. 2007) experimental dependences $C_D$ on $U_{10}$. Stress that these calculations are not part of the present modeling, but only serve to illustrate the influence of possible deviations in measurements of the coverage coefficients on the results of the present modeling.

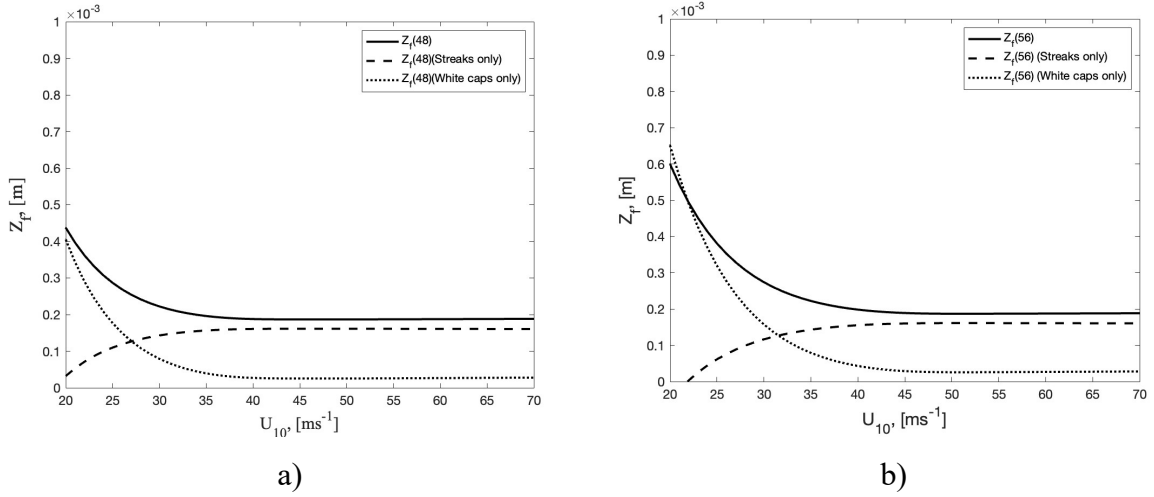

Fig. 4: Dependence of $Z_f$ on $U_{10}$. a) $U_{10}^{(S)} = 48\ [ms^{-1}]$, b) $U_{10}^{(S)} = 56\ [ms^{-1}]$.
Solid line represents the current model, where $Z_f = (\alpha_{wc} Z_{D,wc} + \alpha_{str} Z_{D,str})/\alpha_f$ ;
Dotted line presents the current model, without whitecap contribution, $Z_f = \alpha_{str} Z_{D,str}/\alpha_f$.
Dashed line presents the current model without streak contribution, $Z_f = \alpha_{wc} Z_{D,wc}/\alpha_f$.

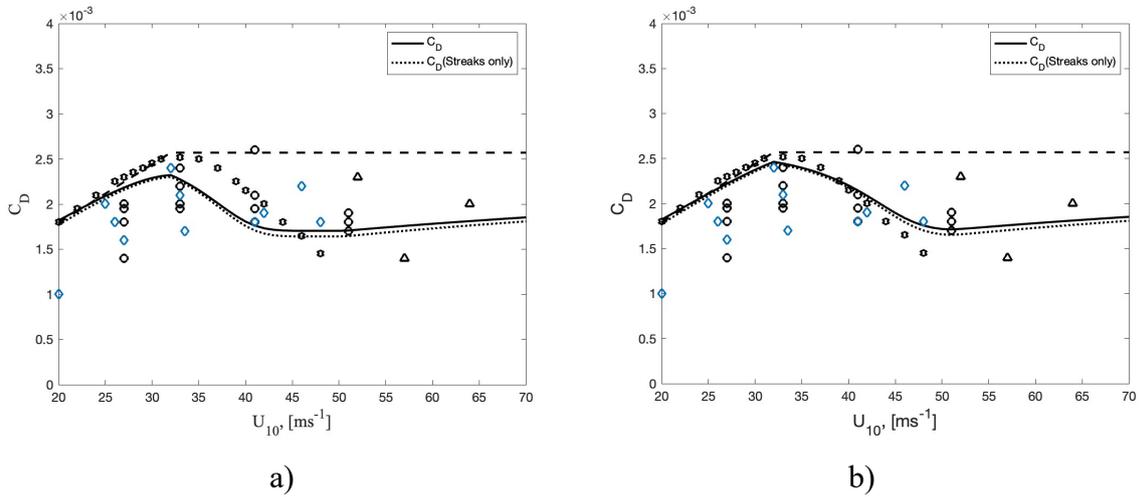

Fig. 5: Dependence of $C_D$ on $U_{10}$. a) $U_{10}^{(S)} = 48\ [ms^{-1}$; b) $U_{10}^{(S)} = 56\ [ms^{-1}]$. Solid line represents the current model for $C_D$; dashed line indicates Large and Ponds 1981 model modified in GS2018 ($C_D = C_{Dw}$); Dotted line shows the current model for $C_D$ without whitecap contribution. Measurement points obtained under open-sea conditions are adopted from the various studies: circles represent Powell et al. 2003, hexagons for Jarosz et al. 2007, diamonds for HPP2012, and triangles for Bell et al. 2012.

*Estimation of the hurricane potential intensity.*

Emanuel (1986) identified that the ratio $C_K/C_D$, is a crucial physical parameter determining hurricane potential intensity. This finding was confirmed through 3D numerical simulations of hurricanes (Bell et al. 2012). Emanuel (1986) provided theoretical estimates for $C_K/C_D$, suggesting that it can vary within the range of 0.75 to 1.5. However, experimental studies indicate a somewhat wider interval of admissible values $C_K/C_D$, namely between 0.5 and 1.5, as shown in Montgomery et al. 2003, Zhang et al. 2008, Bell et al. 2012, Richter et al. 2016. In GS2020, the value of $C_K/C_D$ as a function of $U_{10}$, considering the influence of foam, reached 0.4, which falls outside the aforementioned range. Figure 6 shows that the dependence $C_K/C_D$ on $U_{10}$ as obtained in the current model aligns well with open-sea experimental data. It is observable that the model values of $C_K/C_D$ are greater than 0.5. Note that the value $U_{10}^{(S)}$ has little effect on the behavior of $C_K/C_D$ unlike the behavior of $C_D(U_{10})$.

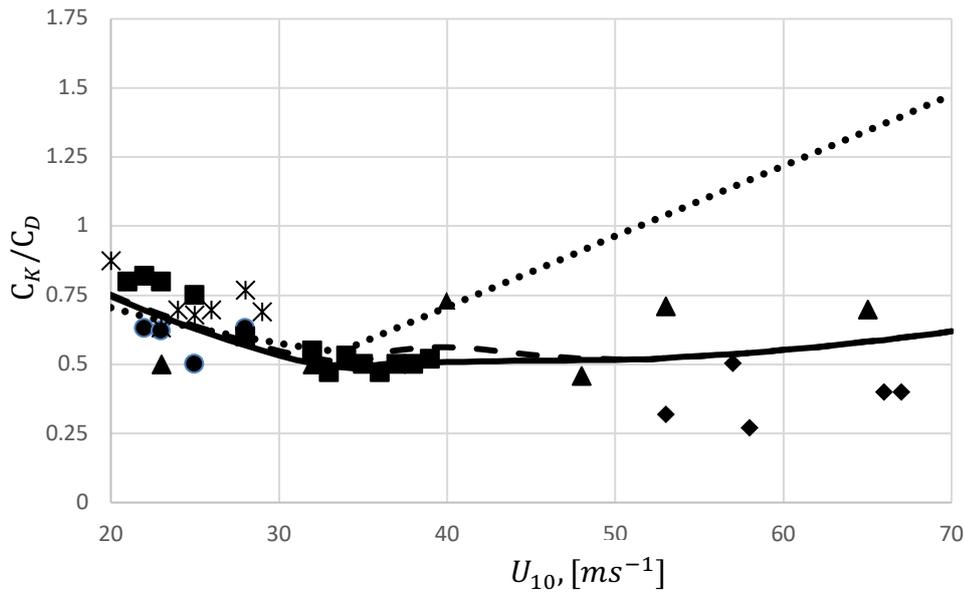

Fig. 6: Dependence of $C_K/C_D$ on $U_{10}$. The solid line represents $U_{10}^{(S)} = 48\ [ms^{-1}]$. The dashed line indicates $U_{10}^{(S)} = 54\ [ms^{-1}]$. The dotted line corresponds to foam-free laboratory experiments by Komori et al. 2018. Measurement points obtained under open-sea conditions are adopted from various studies: circles for Zhang et al. 2008, triangles for Richter et al. 2016, diamonds for Bell et al. 2012, stars for Sergeev et al. 2017, and squares for Haus et al. 2010.

## 4. Conclusions

This paper continues our studies of the foam in the slip layer sandwiched between the atmosphere and the ocean generated by breaking surface waves under hurricane conditions, focusing on the foam's role in the momentum and enthalpy transfer across the layer. The current approach employs a parametrization of the momentum transfer coefficient, $C_D(U_{10})$, in which the effective roughness length, $Z_D(U_{10})$, is split into two roughness lengths for the foam-free- and the foam - covered areas, $Z_{D,w}$ and $Z_{D,f}$, weighted by the coverage coefficients $\alpha_w(U_{10})$ and $\alpha_f(U_{10})$, respectively. Building on our prior studies, we employ laboratory data for foam-free covered areas, which describe all effects of the slip layer except for foam. The fractional foam roughness $Z_{D,f}$ is further split into two sub-fractional roughness lengths, $Z_{D,wc}$ and $Z_{D,str}$, for the whitecap-covered and streak-covered ocean areas, weighted by the coverage coefficients, $\alpha_{wc}(U_{10})$ and $\alpha_{str}(U_{10})$. Developing the idea proposed in GS2016, and considering the characteristics of breaking surface waves, the sub-fractional roughness lengths are identified as geometrical radii of the bubbles in whitecaps and streaks, namely, $Z_{D,wc} = R_{ws}$ and $Z_{D,str} = R_{str}$, which are independent of $U_{10}$. Meanwhile, $Z_{D,f}$ varies with $U_{10}$ via $\alpha_{wc}(U_{10})$ and $\alpha_{str}(U_{10})$, highlighting the significance of whitecaps and streaks at lower and higher speeds, respectively.

Specifically, the present model elucidates how foam generation influences the non-monotonic behavior of $C_d(U_{10})$ under hurricane conditions. At wind speeds below 25 $[ms^{-1}]$, foam roughness is primarily influenced by whitecaps, as shown in Figure 4. At these conditions, the foam coverage area remains small ($\alpha_f(U_{10}) < 0.05$), and the behavior of $C_D$ is mainly governed by $C_{D,w}$. As the wind speed increases, the ocean surface becomes increasingly covered with foam. The percentage of surface covering by whitecaps does not change and is approximately 5%. The roughness begins to be determined by the roughness of the streaks, which becomes dominant at speeds above 35 m/s, when the $C_D$ reaches its maximum. As the wind speed increases further, $\alpha_f(U_{10})$ increases, resulting in almost complete coverage of the surface by foam and a decrease in $C_D$. Within the framework of our model, the dependence of the $C_K/C_D$ on $U_{10}$, a key factor in determining hurricane potential intensity, aligns well with experimental data and falls within the range predicted by prior theoretical and experimental studies.

Summarizing, note the total independence of the input data—foam bubble-size spectra and foam coverage coefficients—and the output data for the momentum and heat transfer coefficients between the ocean and atmosphere in hurricane conditions. A crucial aspect of the present surface-drag modeling, represented by the effective roughness length, is the treatment of the fractional roughness lengths, $Z_{D,ws}$ and $Z_{D,str}$, as the effective radii of foam bubbles in whitecaps, $R_{wc}$, and streaks, $R_{str}$. These lengths are independent of wind speed $U_{10}$, and are entirely governed by the surface-wave breaking. However, the total effective foam roughness length, $Z_{D,f}$, is dependent on the wind speed $U_{10}$ through the fractional coverage coefficients of whitecaps $\alpha_{wc}(U_{10})$ and streaks $\alpha_{str}(U_{10})$. Meanwhile, the size distribution spectra of foam bubbles are well established through numerous experimental studies, the less explored data for foam coverage coefficients (for streaks and whitecaps) can be more straightforwardly investigated, even under hurricane conditions. Although the present approach notably omits consideration of water properties like salinity, temperature, and biological content, which could influence the model's input and output, its applicability is validated by its alignment with trusted experimental data. The innovation of this approach stems from its use of well-established experimental data on foam bubble size distribution spectra, a novel application in this research context.